\documentclass[aps,twocolumn]{revtex4-2}
\usepackage{graphicx}
\usepackage{hyperref}
\usepackage{amssymb}
\usepackage{bm}
\usepackage{color}

\newcommand{\be}{\begin{eqnarray}}
\newcommand{\ee}{\end{eqnarray}}
\newcommand{\no}{\nonumber}

\begin{document}
\title{Dynamical quantum phase transition, metastable state, and dimensionality reduction: 
Krylov analysis of fully-connected spin models}

\author{Kazutaka Takahashi 
\href{https://orcid.org/0000-0001-7321-2571} {\includegraphics[scale=0.05]{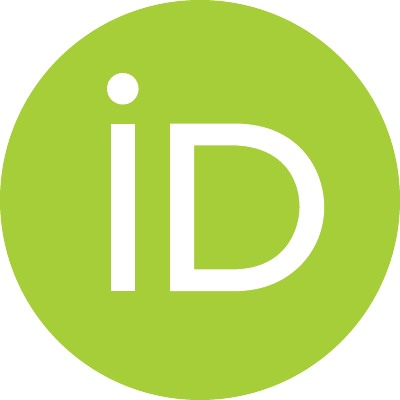}}}
\affiliation{Department of Physics and Materials Science, University of Luxembourg, L-1511 Luxembourg, Luxembourg}
\affiliation{Department of Physics Engineering, Faculty of Engineering, Mie University, Mie 514–8507, Japan}

\begin{abstract}
We study quenched dynamics of fully-connected spin models.
The system is prepared in a ground state of the initial Hamiltonian and 
the Hamiltonian is suddenly changed to a different form.
We apply the Krylov subspace method to map the system onto 
an effective tridiagonal Hamiltonian.
The state is confined in a potential well and is time-evolved by nonuniform hoppings.
The dynamical singularities for the survival probability can occur 
when the state is reflected from a potential barrier.
Although we do not observe any singularity in the spread complexity,
we find that the entropy exhibits small dips at the singular times.
We find that the presence of metastable state affects long-time behavior of 
the spread complexity, and physical observables.
We also observe a reduction of the state-space dimension
when the Hamiltonian reduces to a classical form.
\end{abstract}

\date{\today}

\maketitle

\section{Introduction}

Nonequilibrium dynamics in quantum many-body systems has 
emerged as a central theme in quantum algorithms
and condensed matter 
physics~\cite{Calabrese06,Polkovnikov11,Eisert14,Mitra18}. 
In particular, quantum quenches where a system is prepared 
in an initial state and then subjected to a sudden change 
in the Hamiltonian, serve as a fundamental protocol 
for studying nonequilibrium physics. 
The response of a system to such abrupt changes not only reveals
fundamental properties of quantum many-body systems but also 
provides crucial insights into quantum information propagation 
and thermalization mechanisms.

It is known that a certain kind of quenched systems exhibits
dynamical quantum phase transitions 
(DQPTs)~\cite{Heyl13,Heyl14,Heyl15,Jurcevic17,Heyl18}. 
When we consider quenches across an equilibrium quantum phase transition, 
the rate function of the survival probability as a function of time 
shows nonanalytic behavior at the thermodynamic limit. 
Various quench patterns lead to various behaviors in real-time evolutions
that cannot be seen in the corresponding equilibrium system.
Some of the properties are described by the 
analytic studies of specific exactly solvable models.

In this study, we describe a quenched system with respect to 
the Krylov subspace method~\cite{Viswanath94,Nandy25}.
The quenched dynamics is set by specifying the Hamiltonian 
and the initial state.
Since this setting is completely the same as that 
in the Krylov subspace method, the application of the method 
is a reasonable strategy.
The advantage of the Krylov subspace method
is that any system can be mapped onto a one-dimensional 
hopping model.
The tridiagonal form of the effective Hamiltonian reflects 
the initial setting of the quench protocol.
Then, it would be interesting to describe the DQPT 
with respect to the Krylov terminology.
The fundamental question to be asked is whether 
there is any advantage using
the Krylov subspace method for the description of the DQPT.
The method has attracted renewed interests 
for describing universal properties of operator 
growth~\cite{Perker19,Caputa22}.
To quantify the concept of operator complexity, 
the authors in Ref.~\cite{Perker19} introduced the Krylov complexity 
measuring how far the time-evolved Heisenberg operator is from 
the initial operator.
The same consideration is applied to the time-evolved state and 
we can define the spread complexity
to quantify the state spreading~\cite{Balasubramanian22}.
Is the onset of the DQPT reflected to the spread complexity and/or 
some other quantities defined in Krylov space?

As a simple model showing the DQPT, 
we exploit a fully-connected spin Hamiltonian.
The quenched Lipkin--Meshkov--Glick (LMG) model~\cite{Lipkin65} is 
one of the models that exhibit the DQPT~\cite{Zunkovic16, Obuchi17}.
As shown in Ref.~\cite{Obuchi17}, the singularities can be best seen in
systems with a bias field that breaks spin-reflection symmetry.
Although there exist several preceding studies of the LMG model 
by the Krylov subspace method~\cite{Bhattacharjee22, Afrasiar23, Bento24}, 
and by the non-Krylov complexity analysis~\cite{Pal23}, 
the bias field was not introduced there.
Here, we analyze the LMG model with the bias field.
The main aim of this study is to see how 
the DQPT is described in Krylov space.
We find that, 
in the picture of one-dimensional Krylov lattice, 
the DQPT can occur when the state is reflected from a potential barrier.
Since the hopping amplitude is nonuniform, 
this one-dimensional picture incorporates nontrivial effects.
Similar Krylov studies were recently 
done for Ising models~\cite{Bhattacharya24,Guerra25prb,Shirokov25}.

In principle, the DQPT occurs only in the thermodynamic limit.
Although they appear repeatedly in the time sequence of the survival probability,
the amplitude typically shows decaying behavior, which leads to smearing of singularities.
It is generally difficult to know the long-time behavior of the system,
both theoretically and experimentally.
Theoretically, the long-time behavior is strongly dependent on the system size 
and it is hard to know the thermodynamic limit.
In this study, we introduce the bias field.
It explicitly breaks spin-reflection symmetry and makes the system 
more in a trivial state.
However, a small bias field produces 
a metastable state, which makes the dynamical behavior nontrivial.
While the metastable state does not affect
the statistical-mechanical properties at the thermodynamic limit,
it does dynamical properties significantly.
We find that the long-time behavior is unstable 
and is sensitive to the parameter choice.

The organization of this paper is as follows.
In Sec.~\ref{sec:modelandmethod} we introduce the model system, 
summarize the known results, and describe our strategy.
Next, in Sec.~\ref{sec:dqpt}, we study dynamical singularities
by using the Krylov subspace method.
We also discuss metastable state in Sec.~\ref{sec:meta} and
dimensionality reduction in Sec.~\ref{sec:dim}.
The last section \ref{sec:conc} is devoted to conclusions.

\section{Krylov subspace method for the LMG model}
\label{sec:modelandmethod}

\subsection{DQPT and Krylov subspace method}

In the standard framework of closed quantum systems,
we prepare a initial state $|\psi_0\rangle$ and 
consider the time evolution
\be
 |\psi(t)\rangle = e^{-iHt}|\psi_0\rangle.
\ee
Here, $H$ represents the Hamiltonian of the system.
When the initial state is not equal to one of the eigenstates of $H$,
the time evolution gives nontrivial states.
In particular, for many-body systems, the initial state can be 
a sum of many eigenstates, which induces nontrivial effects.
We are mainly interested in the survival amplitude 
$\langle\psi_0|\psi(t)\rangle$.
For a typical many-body system with a large value of the system size $N$, 
this quantity is exponentially small and 
it is reasonable to define the rate function 
\be
 f(t) = -\frac{1}{N}\ln|\langle\psi_0|\psi(t)\rangle|.
 \label{rate}
\ee
Then, at the thermodynamic limit $N\to\infty$, 
this function can exhibit singularities for quenches involving 
large changes in parameters~\cite{Heyl13,Heyl14,Heyl15,Heyl18}.
Although the DQPT is likely to occur when the change goes across 
the equilibrium quantum phase transition point, 
the precise value is not necessary equal to that point,
as we show below in our example.

The Krylov subspace method is ideal to treat such systems
as it identifies the minimal subspace of the time evolution.
We set $|K_0\rangle = |\psi_0\rangle$ 
and construct the orthonormal Krylov-basis series
from the three-term recurrence relation 
\be
 |K_{k+1}\rangle b_{k+1}= H|K_k\rangle -|K_k\rangle a_k -|K_{k-1}\rangle b_k, \label{three}
\ee
where $k$ runs as $k=0,1,\dots,d-1$ and 
\be
 && a_k = \langle K_k|H|K_k\rangle, \\
 && b_k = \langle K_{k-1}|H|K_k\rangle.
\ee
We note that $a_k$ is defined for $k=0,1,\dots,d-1$
and $b_k$ is for $k=1,2,\dots,d-1$.
In Eq.~(\ref{three}), We set formally $b_0=0$ for $k=0$ 
and $b_d=0$ for $k=d-1$.
The phase of $|K_k\rangle$ is chosen so that $b_k$ is positive.
The number of the basis $d$ is called Krylov dimension and 
is equal to or smaller than the Hilbert space dimension.

When the time-evolved state is expanded as 
\be
 |\psi(t)\rangle = \sum_{k=0}^{d-1}|K_k\rangle \varphi_k(t), 
\ee
the set of coefficient functions $\{\varphi_k(t)\}_{k=0}^{d-1}$ satisfies 
\be
 i\partial_t\varphi_k(t)= a_k\varphi_k(t)
 +b_{k}\varphi_{k-1}(t)+b_{k+1}\varphi_{k+1}(t). \label{varphit}
\ee
This relation denotes that the state exhibits a one-dimensional spreading motion 
when it is represented in the Krylov space.
To characterize the spreading in the time evolution,
we use the (spread) 
complexity~\cite{Perker19,Balasubramanian22} 
\be
 K(t)=\sum_{k=0}^{d-1} k|\varphi_k(t)|^2, \label{k}
\ee
and the entropy~\cite{Barbon19, Fan22}
\be
 S(t)=-\sum_{k=0}^{d-1} |\varphi_k(t)|^2\ln |\varphi_k(t)|^2. \label{s}
\ee

The effective Hamiltonian in the Krylov space takes a tridiagonal form 
and is represented by the Lanczos coefficients.
Each diagonal component $a_k$ represents 
the local potential at discrete site $k$, 
and $b_k$ represents the hopping amplitude between $k-1$ and $k$.
The system is equivalent to 
the discretized system with a local potential and a site-dependent mass.
The state favors smaller $a_k$ and larger $b_k$.

Our aim in this study is to describe 
the DQPTs from the Krylov picture.
However, we note that the survival amplitude is given 
by the zeroth component 
$\varphi_0(t)=\langle K_0|\psi(t)\rangle$.
This does not contribute to $K(t)$
and the contribution to $S(t)$, 
$-|\varphi_0(t)|^2\ln|\varphi_0(t)|^2$, is negligibly small
at the singular points.
It is not obvious how the singularity is described in the one-dimensional picture.

\subsection{Hamiltonian and phase diagram}
\label{sec:phase}

Our spin model is written with respect to the spin operator
$\bm{S}=(S^x,S^y,S^z)$.
The quantum number $\bm{S}^2=S(S+1)$ is conserved and
we take $S=N/2$ with an integer $N$.
We are interested in the large-$N$ behavior.
In the following calculation, to avoid cumbersome notation, 
we assume $N$ is an even number.

The LMG Hamiltonian is written as 
\be
 H= -2J\left[\frac{1}{N}(S^z)^2+hS^z+gS^x\right]. \label{Ham}
\ee
and is parametrized by $(J,h,g)$.
We take $J>0$, $h\ge 0$, and $g\ge 0$.
The scale of the system is measured in units of $J$ and
all results are represented as functions of dimensionless parameters
$h$ and $g$.
The longitudinal bias field $h$ plays the role of symmetry breaking
and the transverse field $g$ introduces
quantum fluctuation effects.
Since the spin operator is interpreted as the sum of $1/2$-spins,
$\bm{S}=\sum_{i=1}^N\bm{\sigma}_i/2$, 
this model is equivalent to the fully-connected quantum Ising model.

As a quenched time evolution, we set that
the initial state $|\psi_0\rangle$
is the ground state at the Hamiltonian with $g\to\infty$.
When we define the eigenstates of $S^x$ as 
\be
 S^x|m\rangle_x = m|m\rangle_x, \label{mx}
\ee
the eigenvalues take $m=-S,-(S-1),\dots,S-1,S$ and 
the initial state is given by 
\be
 |\psi_0\rangle = |S\rangle_x. \label{initial}
\ee
In the $S^x$-eigenstate basis, the Hamiltonian is represented in a pentadiagonal form as 
\be
 \frac{H}{NJ} &=& 
 -\sum_{k=0}^{N} \left[\frac{C_k^2+C_{k+1}^2}{2N^2}+\left(1-\frac{2k}{N}\right)g\right]
 \no\\ && \times
 |S-k\rangle_x{}_x\langle S-k| \no\\
 && -\sum_{k=1}^{N}\frac{hC_k}{N} 
 \no\\ && \times
 \left(|S-k+1\rangle_x{}_x\langle S-k|+|S-k\rangle_x{}_x\langle S-k+1|\right) \no\\
 && -\sum_{k=2}^{N}\frac{C_{k-1}C_{k}}{2N^2}
 \no\\ && \times
 \left(|S-k+2\rangle_x{}_x\langle S-k|+|S-k\rangle_x{}_x\langle S-k+2|\right), \no\\
 \label{Hamx}
\ee
where $C_k=\sqrt{k(N+1-k)}$.
By applying the Krylov algorithm, we can transform it to a tridiagonal form,
which is the main task in the following sections.

We note that the procedure is greatly simplified at $h=0$.
In that case, only $|m\rangle_x$ with $m=-S, -(S-2), \dots, S-2, S$ 
contribute to the time evolution and they give the Krylov basis set.
The original Hamiltonian is in a tridiagonal form
and the Krylov dimension is given by $d=S+1=\frac{N}{2}+1$ 
which is almost half of the Hilbert space dimension $2S+1=N+1$.
Since the dynamical singularities on the rate function 
in Eq.~(\ref{rate}) are clearly observed 
for nonzero values of $h$~\cite{Obuchi17} and we can find 
preceding Krylov studies at $h=0$~\cite{Bhattacharjee22,Afrasiar23,Bento24}, 
we basically consider $h>0$ in the following calculations.

\begin{figure}[t]
\centering\includegraphics[width=1.\columnwidth]{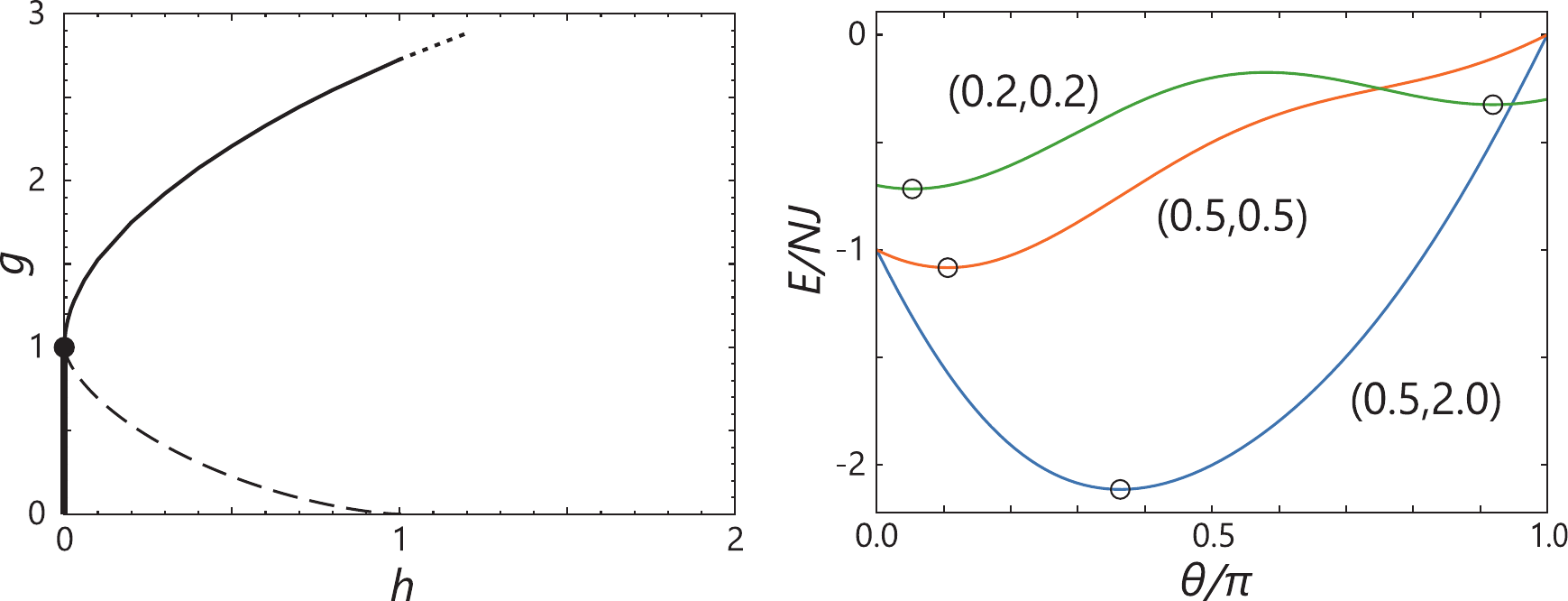}
\caption{Left: Phase diagram of the quenched LMG model.
The phase boundary is shown by the bold line
and the critical point by the dot.
The solid line obtained from a semiclassical analysis in Ref.~\cite{Obuchi17}
represents the boundary on the DQPTs.
Below the sold line, we observe dynamical singularities of $f(t)$.
The metastable state exists below the dashed line.
Right: Equation (\ref{e}) for several values of $(h,g)$.
The circle marks denote local minimum points.
}
\label{fig-pd}
\end{figure}

A possible behavior is roughly inferred from the equilibrium statistical properties
of the Hamiltonian in the thermodynamic limit $N\to\infty$.
We show the phase diagram in the left panel of Fig.~\ref{fig-pd}.
At the limit, the system is described semiclassically
and the ground state is evaluated by parameterizing the spin as 
$\bm{S}=\frac{N}{2}(\sin\theta,0,\cos\theta)$
with $0\le\theta\le\pi$.
The ground-state energy is written as
$E_\mathrm{gs}=\min_\theta E(\theta)$ where
\be
 E(\theta) = -NJ\left(
 \frac{1}{2}\cos^2\theta+h\cos\theta+g\sin\theta\right). \label{e}
\ee
We show the function for several values of $(h,g)$ in the right panel of Fig.~\ref{fig-pd}.
When the bias field $h$ is absent,
$E(\theta)$ is minimized at $\theta=\pi/2$ at $g\ge 1$
and at $\theta=\arcsin g$ at $g\le 1$.
The latter has two possible solutions of $\theta$,
showing the spin-reflection symmetry.
The point $(h,g)=(0,1)$ is identified as the critical point.
For nonzero values of $h$,
$\arg\min_\theta E(\theta)$ uniquely exists and
we observe no sharp transition.

When $h$ and $g$ are small enough, we observe a local minimum
in addition to the global minimum.
The local minimum represents the metastable state and exists when
\be
 g< (1-h^{2/3})^{3/2},
\ee
with $0<h<1$.
The spinodal line representing the boundary is shown 
in the left panel of Fig.~\ref{fig-pd}.

\begin{figure}[t]
\centering\includegraphics[width=1.\columnwidth]{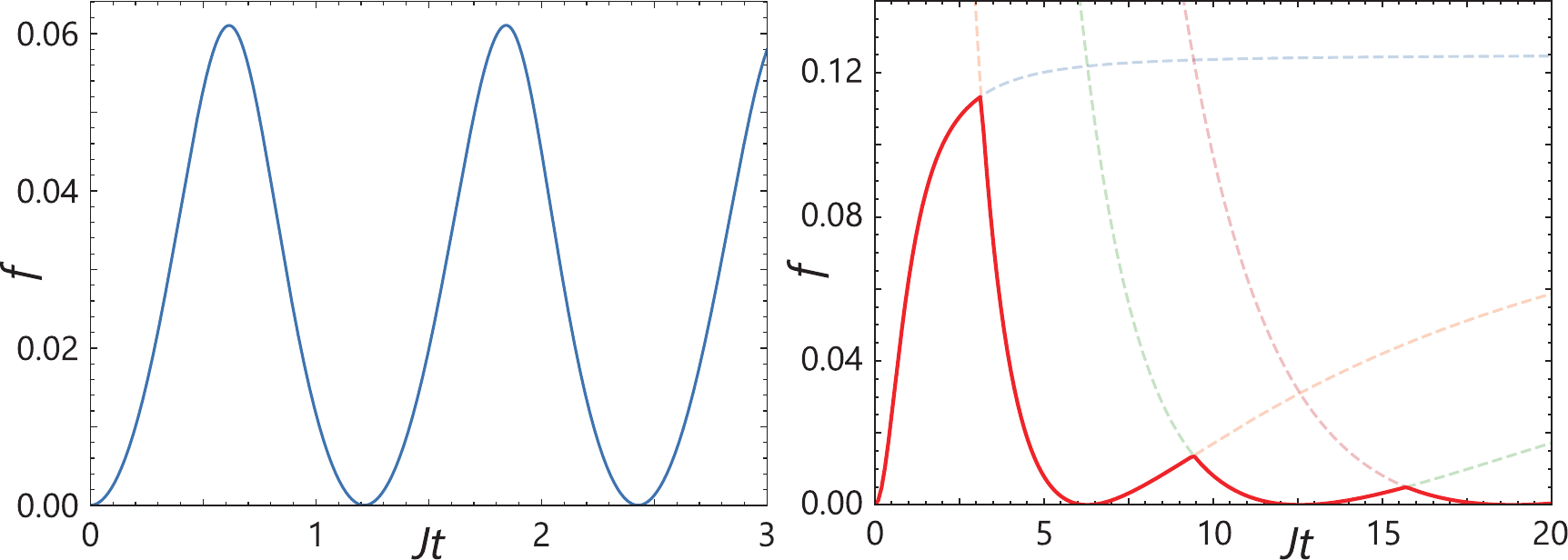}
\caption{The rate function $f(t)$ for the quenched LMG model.
The result on the left panel is obtained numerically
with the parameters $(h,g)=(0.5,3.0)$ and $N=400$.
The right panel represents Eq.~(\ref{rate-exact}) for $(h,g)=(0.5,0.0)$.
}
\label{fig-dqpt}
\end{figure}

A semiclassical analysis of the survival amplitude was closely discussed 
in Ref.~\cite{Obuchi17}.
It was shown that the rate function $f(t)$ at $N\to\infty$ 
has singular points when $g$ is small.
As a noteworthy result, the rate function at $g=0$ is evaluated as
\be
 \lim_{N\to\infty}f(t)=\min_{n\in\mathbb{Z}}\frac{(hJt-\pi n)^2}{2[1+(Jt)^2]}. 
 \label{rate-exact}
\ee
This result is obtained not from the above-mentioned semiclassical analysis 
but from a simple saddle-point analysis justified at $N\to 0$.
Since the Hamiltonian at $g=0$ is diagonal in the $S^z$-eigenstate basis, 
we can utilize the standard statistical-mechanical analysis for mean-field systems. 
By using the Poisson summation formula, we replace the sum over the basis
to an integral form and use the saddle point to evaluate the integral~\cite{Obuchi17}.
We show typical behavior of the rate function in Fig.~\ref{fig-dqpt}.

Thus, we conclude the phase diagram of the quenched LMG model
in the left panel of Fig.~\ref{fig-pd}.
We note that our definition of the dynamical singularity is applied
to the survival amplitude.
The introduction of the symmetry-breaking $h$ 
implies that we do not observe singularities for the order parameter, 
the expectation value of $S^z$.

\subsection{Representation in $S^z$-eigenstate basis}
\label{sec:zbasis}

Although the $S^x$-eigenstate basis is convenient for the initial state, 
it is not for the Hamiltonian.
We can switch to the $S^z$-eigenstate basis representation to write 
\be
 \frac{H}{NJ} &=& -\sum_{k=0}^{N}\left[2\left(\frac{k}{N}-\frac{1+h}{2}\right)^2
 -\frac{h^2}{2}\right]|S-k\rangle_z{}_z\langle S-k| \no\\
 && -\sum_{k=1}^{N}g\sqrt{\frac{k}{N}\left(1-\frac{k}{N}\right)} \no\\
 && \times\left(|S-k+1\rangle_z{}_z\langle S-k|+|S-k\rangle_z{}_z\langle S-k+1|\right), \no\\
 \label{Hamz}
\ee
where we use the $S^z$ basis $|m\rangle_z$ with $m=\frac{N}{2}-k$.
In the same basis, the initial state in Eq.~(\ref{initial}) is written as 
\be
 |\psi_0\rangle &=& \sum_{k=0}^N\left(\frac{1}{2}\right)^{N/2}
 \sqrt{\left(\begin{array}{c} N \\ k \end{array}\right)}|S-k\rangle_z \no\\
 &\sim& \sum_{k=0}^N\left(\frac{2}{\pi N}\right)^{1/4}
 \exp\left[-N\left(\frac{k}{N}-\frac{1}{2}\right)^2\right]|S-k\rangle_z. 
 \no\\ \label{psi0z}
\ee
In the second line, we use the Stirling's approximation which is justified at large $N$.
Since the initial state is localized in the $x$-basis, 
it is extended in the $z$-basis.
However, when we take the large-$N$ limit, the state is localized around the point $m=0$
with a width proportional to $\sqrt{N}$.

Thus, when we consider large values of $N$, we observe spreading 
of the zero-magnetization state to nonzero states.
The first term of Eq.~(\ref{Hamz}) plays the role of potential and the second term
represents nearest-neighbor hopping.
For large $g$, the hopping term is the dominant contribution.
Since the hopping amplitude is maximum around the initial state, 
the state oscillates around the initial state.
When $g$ takes a smaller value, the potential term enhances the spreading 
toward the positive-magnetization direction.
When it reaches the state $k=0$, we observe a reflection,
giving rise to a nontrivial interference of the wave function.
We also see that the spreading toward the negative direction is more complicated.
Although the increasing potential prevents the state from spreading, 
the potential produces a local minimum at $k=N$ when $h\le 1$.
It represents the metastable state and we expect a nontrivial behavior.

Thus, by using the $S^z$-eigenstate representation, 
we can develop a qualitative picture of the time evolution.
To make the picture more quantitative, 
we apply the Krylov algorithm to our Hamiltonian.

\section{Dynamical quantum phase transitions}
\label{sec:dqpt}

We numerically calculate the Krylov basis and the Lanczos coefficients.
Although the three-term recurrence in Eq.~(\ref{three})
guarantees the orthonormality $\langle K_m|K_n\rangle=\delta_{m,n}$,
numerical calculations give an accumulation of errors. 
We use the full orthogonalization procedure~\cite{Rabinovici21},
which means that we use the update  
$|K_n\rangle\to |K_n\rangle
-\sum_{m=0}^{n-1}|K_m\rangle\langle K_m|K_n\rangle$ 
and the normalization for each step $n$.

By using the obtained Lanczos coefficients, 
we solve the time evolution in Eq.~(\ref{varphit}) 
to calculate the complexity $K(t)$ and the entropy $S(t)$.
We also calculate the expectation values of $S^z$ and $S^x$
from the time evolution $|\psi(t)\rangle$ without the Krylov algorithm. 

\subsection{Lanczos coefficients}

\begin{figure}[t]
\centering\includegraphics[width=1.\columnwidth]{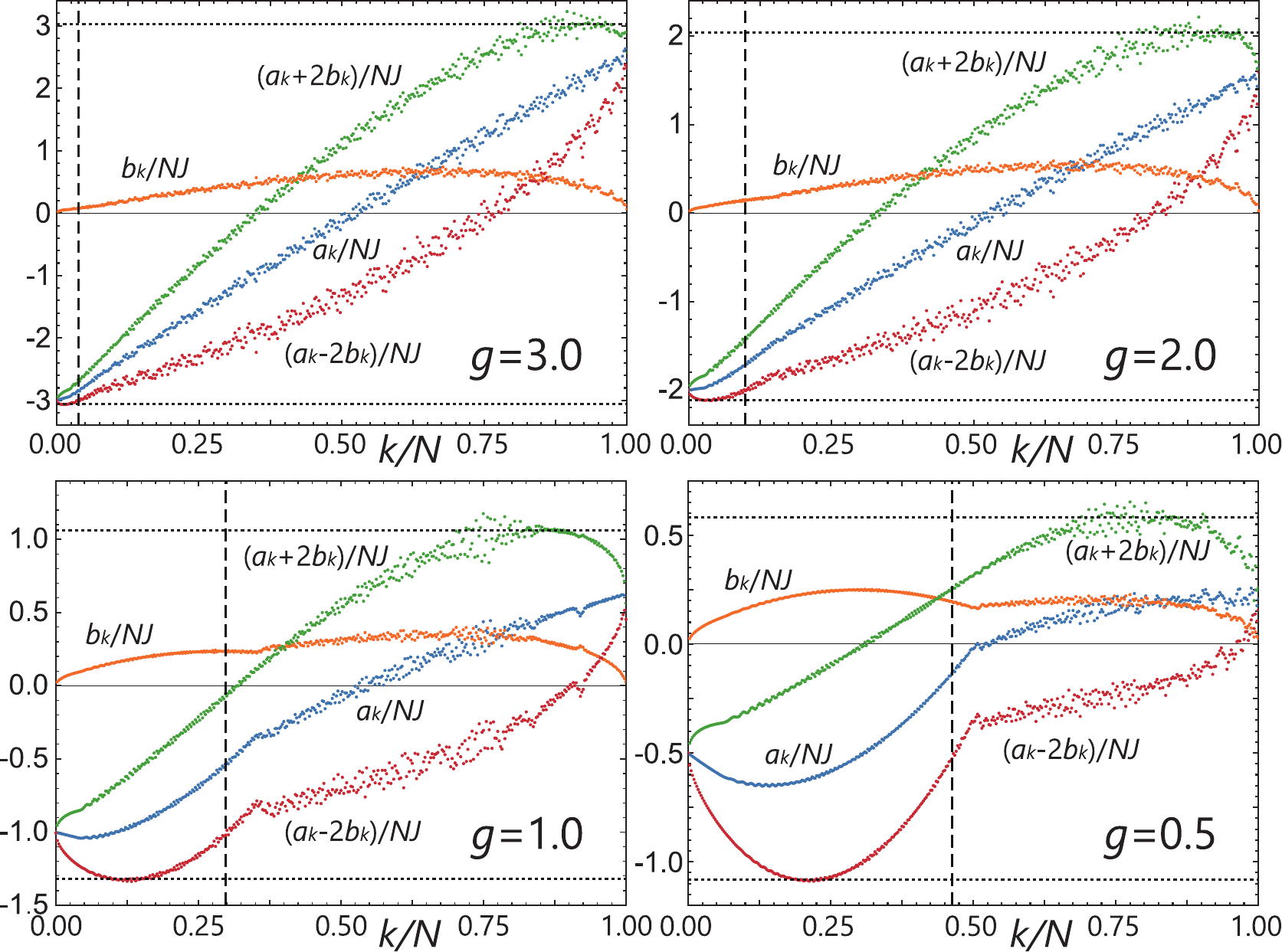}
\caption{
The Lanczos coefficients $a_k$ and $b_k$ for $N=400$, $h=0.5$, 
and $g=3.0, 2.0, 1.0, 0.5$.
The horizontal dotted lines in each panel represent 
the maximum and minimum eigenvalues of the Hamiltonian, in units of $NJ$.
The vertical dashed line represents 
the maximum value of the complexity, $\max_t K(t)/N$.
The complexity is shown in Fig.~\ref{fig-ks}.
}
\label{fig-ab}
\end{figure}
\begin{figure}[t]
\centering\includegraphics[width=1.\columnwidth]{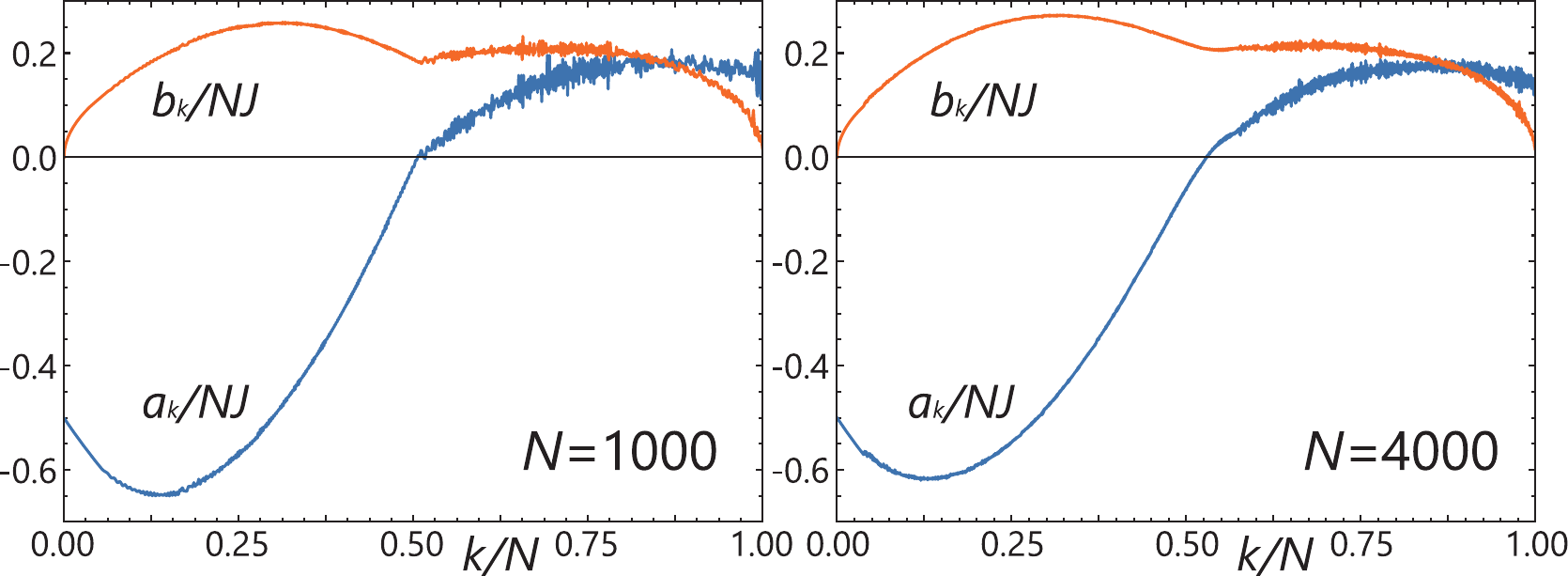}
\caption{
The size dependence of the Lanczos coefficients for $(h,g)=(0.5,0.5)$.
We take $N=1000$ for the left panel and 4000 for the right.
}
\label{fig-abn}
\end{figure}

In Fig.~\ref{fig-ab}, 
we plot the Lanczos coefficients for the system size $N=400$.
We fix $h=0.5$ and take several values for $g$.
When $g$ is large enough and the state has no DQPT, 
$a_k$ grows almost linearly and 
$b_k$ is almost constant compared to the grow scale of $a_k$.
The slope of $a_k$ is large enough and the time-evolved state 
cannot go far from the initial state.
This result is consistent with the picture developed in 
Sec.~\ref{sec:zbasis}.

When we take a smaller $g$, $a_k$ takes the minimum at $k\ne 0$ and 
$b_k$ has an inverted parabola-like form 
with the maximum around the middle of the index range.
We also observe two-domain structures both in $a_k$ and $b_k$ 
and growings of the first domain for decreasing $g$.
The slope of $a_k$ for small $k$ is understood from 
the analytic evaluation 
\be
 a_1-a_0 = 2J\left(g-\frac{3}{2}\right)+O(N^{-1}). \label{slope}
\ee
We demonstrate the derivation in the Appendix.
The nonzero minimum point appears at $g<3/2$.
The range width of $a_k$, $\max_k a_k-\min_k a_k$, becomes smaller 
as $g$ decreases and the time-evolved state can reach 
higher Krylov-basis states.
As we see in Fig.~\ref{fig-abn}, the two-domain structure is 
preserved for larger values of $N$.
A similar structure was numerically observed in the same model 
with a time-dependent modulation in Ref.~\cite{Takahashi25}.

The asymptotic form at $N\to\infty$ is obtained by parametrizing 
the Hamiltonian by indices $x_k=k/N$ with $0\le x_k\le 1$.
By using Eq.~(\ref{Hamx}), we can find 
\be
 \frac{H}{NJ} &\sim& 
 -\sum_{k=0}^{N} \left[x_k(1-x_k)+g(1-2x_k)\right]|S-k\rangle_x{}_x\langle S-k| \no\\
 && -\sum_{k=1}^{N}h\sqrt{x_k(1-x_k)} 
 \no\\ && \times
 \left(|S-k+1\rangle_x{}_x\langle S-k|+|S-k\rangle_x{}_x\langle S-k+1|\right) \no\\
 && -\sum_{k=2}^{N}\frac{1}{2}x_k(1-x_k)
 \no\\ && \times
 \left(|S-k+2\rangle_x{}_x\langle S-k|+|S-k\rangle_x{}_x\langle S-k+2|\right). \no\\
\ee

When $a_k$ and $b_k$ change smoothly as functions of the index $k$, 
we can introduce the continuum representation as 
$(a_k, b_k) \to (a(x_k),b(x_k))$ at $N\to\infty$.
The density of state for the Hamiltonian
is written as~\cite{Balasubramanian23} 
\be
 \frac{\mathrm{Tr}\,\delta(E-H)}{\mathrm{Tr}\,1} \sim
 \int_0^1 \frac{dx}{\pi}\,\frac{\Theta(4b^2(x)-(E-a(x))^2)}{\sqrt{4b^2(x)-(E-a(x))^2}}, 
\ee
where $\Theta$ represents the step function.
This representation denotes that the eigenvalues distribute in 
the range $[\min_k(a_k-2b_k),\max_k(a_k+2b_k)]$.
The result in Fig.~\ref{fig-ab} supports this property.
We discuss in the previous section that 
$a_k$ corresponds to the local potential.
To put it more accurately, 
subtracting the contribution from the kinetic energy,
we can identify $a_k-2b_k$ as the local potential.

In Fig.~\ref{fig-ab}, we also denote 
the maximum value of the complexity to be discussed in the following.
As we see in the figure, the maximum value 
can be estimated from the local potential $a_k-2b_k$.
The state is confined in a potential well and oscillates between 
$k=0$ and $k=\max_t K(t)$ where $a_k-2b_k$ takes an identical value.

\subsection{Complexity and entropy}

\begin{figure}[t]
\centering\includegraphics[width=1.\columnwidth]{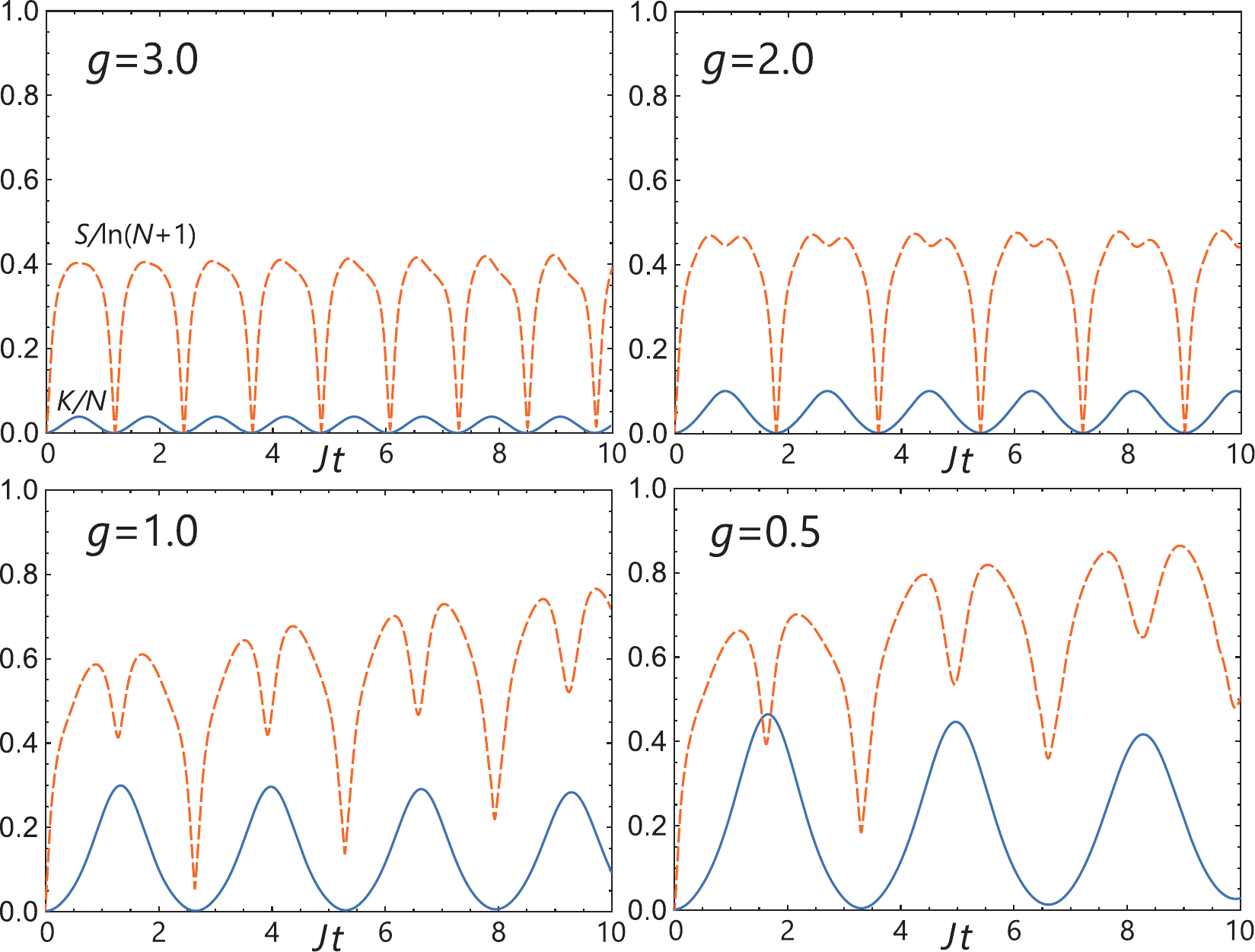}
\caption{
The complexity $K(t)$ (blue solid lines) and 
the entropy $S(t)$ (red dashed lines)
for $N=400$, $h=0.5$, and $g=3.0, 2.0, 1.0, 0.5$.
For $g\le g_c\sim 2.0$, 
the dynamical singularities 
of the rate function in Eq.~(\ref{rate}) 
are obtained at the peak points of $K(t)$.
}
\label{fig-ks}
\end{figure}
\begin{figure}[t]
\centering\includegraphics[width=1.\columnwidth]{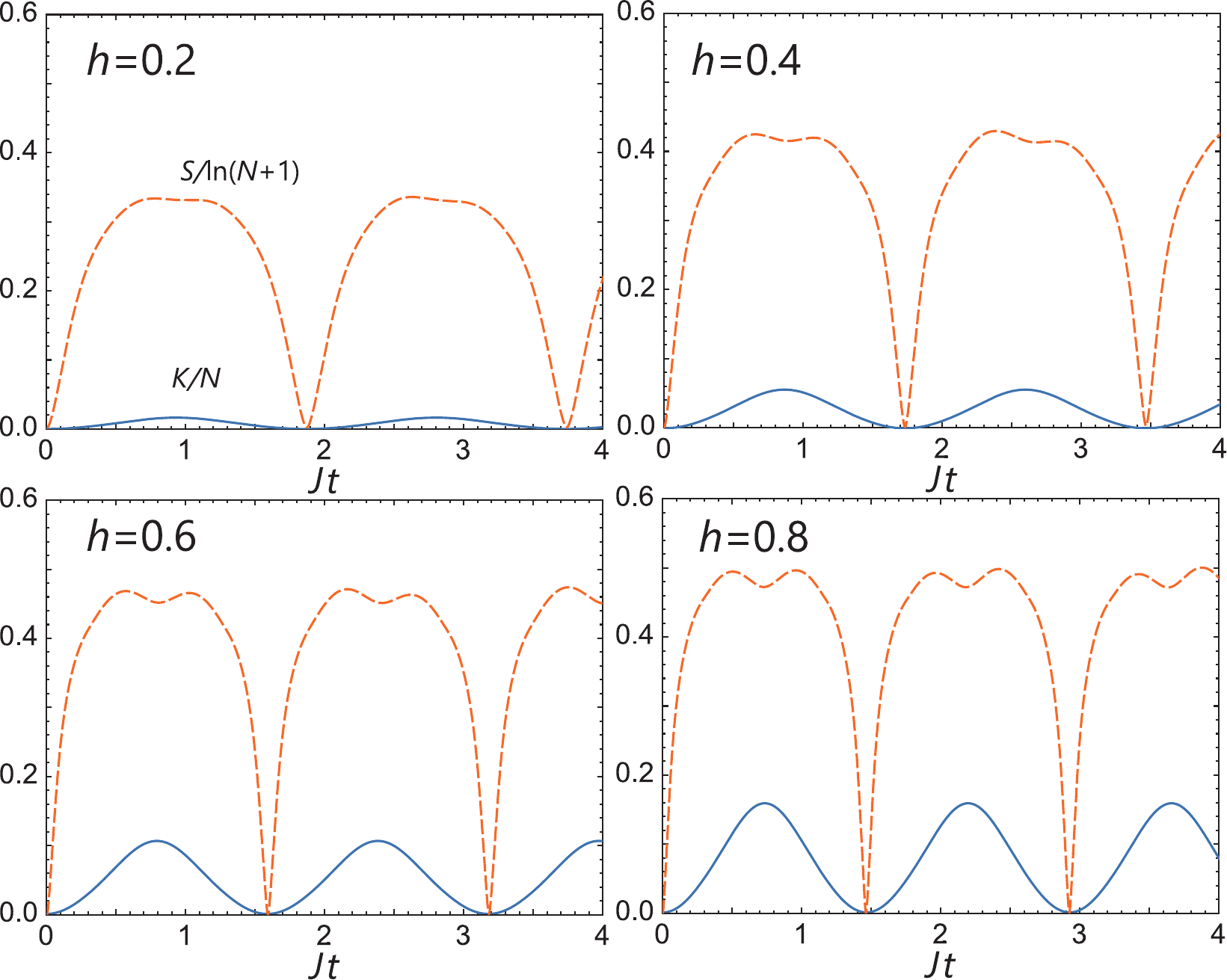}
\caption{
The complexity $K(t)$ (blue solid lines) and 
the entropy $S(t)$ (red dashed lines)
for $N=400$, $g=2.2$, and $h=0.2, 0.4, 0.6, 0.8$.
The dynamical singularities 
are obtained when $h\ge h_c\sim 0.5$.
}
\label{fig-ks2}
\end{figure}
\begin{figure}[t]
\centering\includegraphics[width=1.\columnwidth]{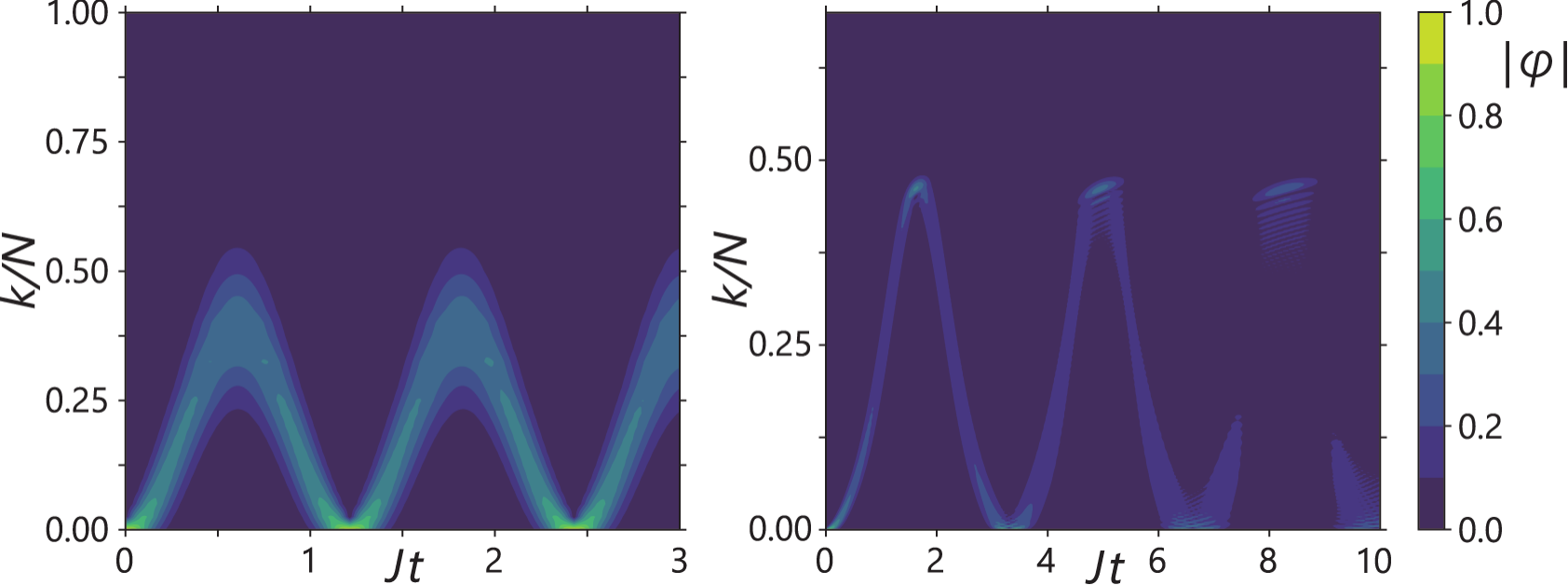}
\caption{
Distributions of $|\varphi_k(t)|$.
We take $(h,g)=(0.5,3.0)$ for the left panel and 
$(0.5,0.5)$ for the right.
The system size is $N=400$.
}
\label{fig-kdist}
\end{figure}
\begin{figure}[t]
\centering\includegraphics[width=1.\columnwidth]{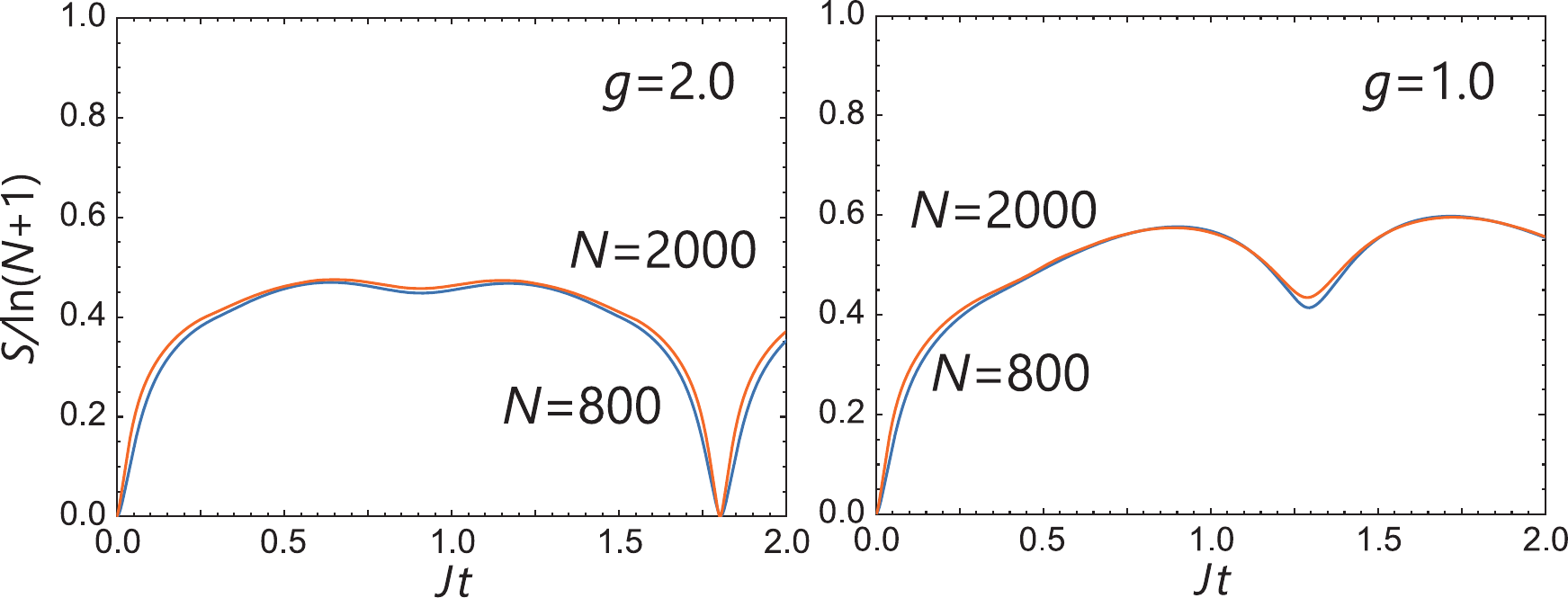}
\caption{
The entropy $S(t)$ for $N=800$ and $2000$.
We take $(h,g)=(0.5,2.0)$ for the left panel 
and $(0.5,1.0)$ for the right.
}
\label{fig-s-n}
\end{figure}

As we see in Fig.~\ref{fig-ab}, 
the Lanczos coefficients with large indices 
are unstable and we observe small oscillations.
However, the unstable fluctuations do not affect 
the actual time evolution 
because the state spreading is basically restricted to lower indices.
We show the complexity $K(t)$ in Figs.~\ref{fig-ks} and \ref{fig-ks2}.
The height of the first peak represents the maximum value of $K(t)$ 
and is denoted in Fig.~\ref{fig-ab}.
As we see in Fig.~\ref{fig-kdist}, the distribution of 
the complexity is a single-peaked function at large $g$.
When $g$ is small, the complexity has broad distributions as $t$ grows.
The result is sensitive to the choice of the system size, and 
it is difficult to know the exact long-time behavior.
In the next section, we study some more details on 
the distributions of the complexity.

When $g$ is small, the dynamical singularities of the rate function 
in Eq.~(\ref{rate}) appear at the peak points of $K(t)$.
It means that the DQPT is obtained when the state 
is reflected by the potential at far-reaching points.
We however find that the complexity does not exhibit any singular behavior.
This is because the complexity is given by the sum of 
many components of $\varphi_k$, 
while the survival probability is obtained only from the zeroth component.
Furthermore, we need take the logarithm 
of the survival probability, as Eq.~(\ref{rate}), to find the singularity.

We also plot the entropy $S(t)$ in Figs.~\ref{fig-ks} and \ref{fig-ks2}.
When $g$ is large enough and no DQPT is observed, 
$S(t)$ shows a similar oscillation as $K(t)$.
The entropy representing an uncertainty of the state 
is maximized when the state reaches a reflection point.
This behavior is changed when $g$ is small.
We observe small dips at the DQPT points.
As we see in Fig.~\ref{fig-s-n}, 
no singular behavior is obtained 
up to considerably large values of $N$.
Since the zeroth-component contribution 
$-|\varphi_0(t)|^2\ln |\varphi_0(t)|^2$ to $S(t)$
is exponentially small in $N$ and is negligible, 
this nonsingular result is reasonable.
Our result shows that 
the DQPT involves a structural change of the entropy.

\subsection{Average of spin operators}

\begin{figure}[t]
\centering\includegraphics[width=1.\columnwidth]{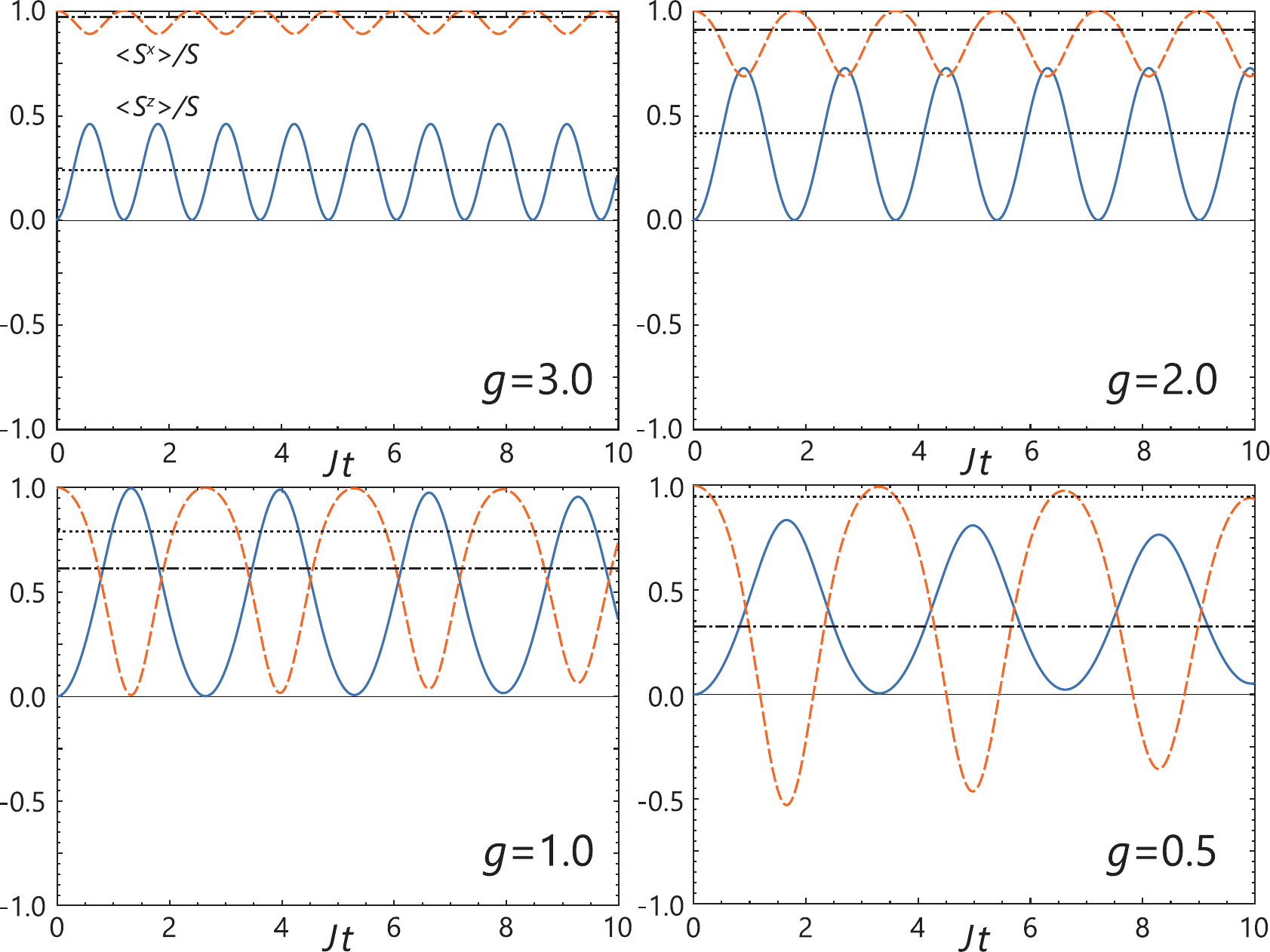}
\caption{
The expectation values of spin operators 
$S^z$ (blue solid lines) and $S^x$ (red dashed lines)
for $N=400$, $h=0.5$, and $g=3.0, 2.0, 1.0, 0.5$.
The dotted line in each panel represents 
the equilibrium value of $S^z$ at the ground state, 
and the dot-dashed line represents the value of $S^x$.
}
\label{fig-spin}
\end{figure}
\begin{figure}[t]
\centering\includegraphics[width=1.\columnwidth]{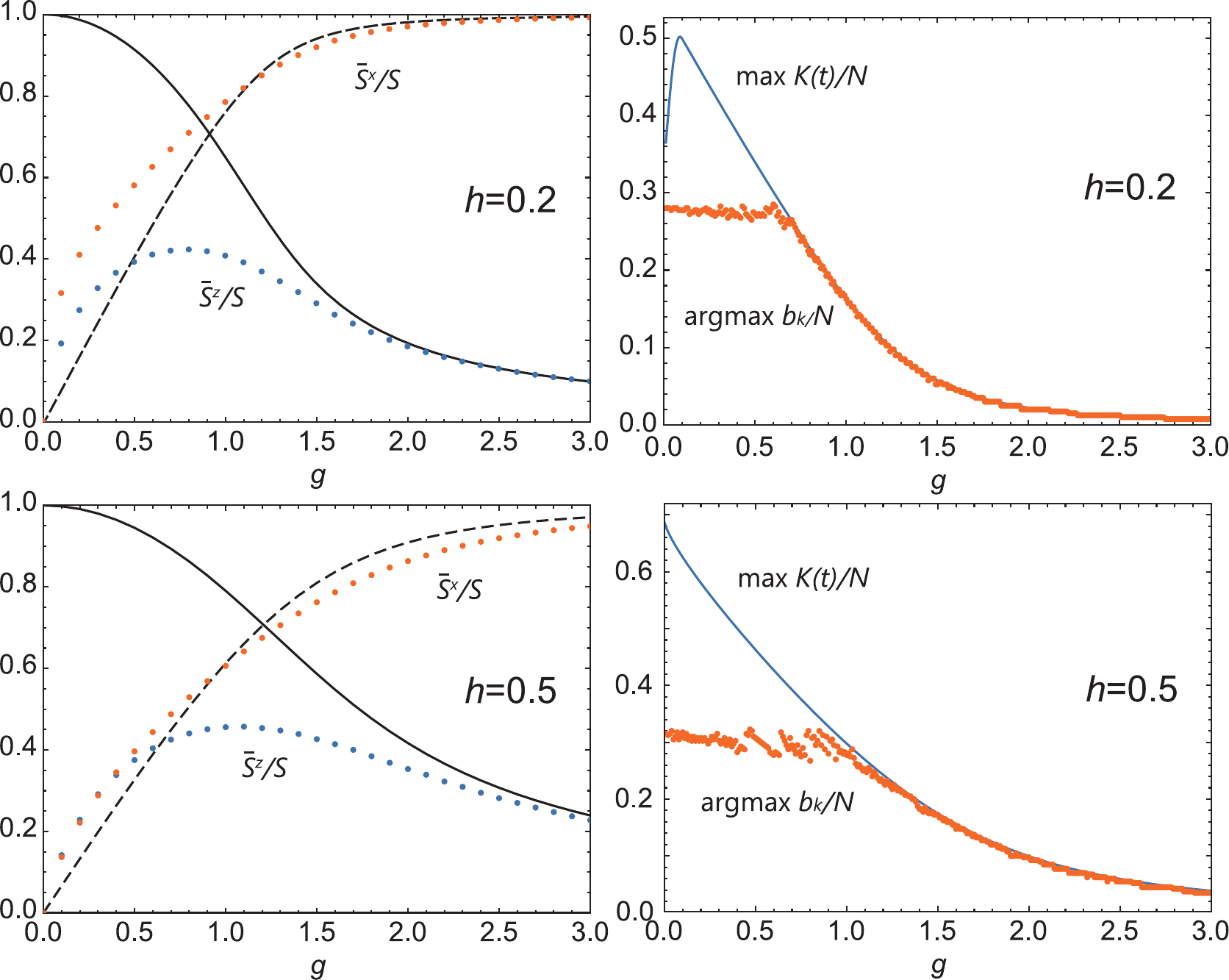}
\caption{
Left: The time-averaged expectation values of $S^z$ 
(blue dots) and $S^x$ (red dots).
We take $h=0.2$ for the upper panel and 
$h=0.5$ for the lower panel.
The system size is $N=400$.
The time average is taken over the range $0\le Jt\le 100$.
The dashed line represents 
the equilibrium value of $S^z$ at the ground state, 
and the dotted line represents the value of $S^x$.
Right: $\max_t K(t)$ (blue solid line) and 
$\mathrm{argmax}_{k\le \max_t K(t)}\, b_k$ (red dots).
}
\label{fig-spinav}
\end{figure}
\begin{figure}[t]
\centering\includegraphics[width=1.\columnwidth]{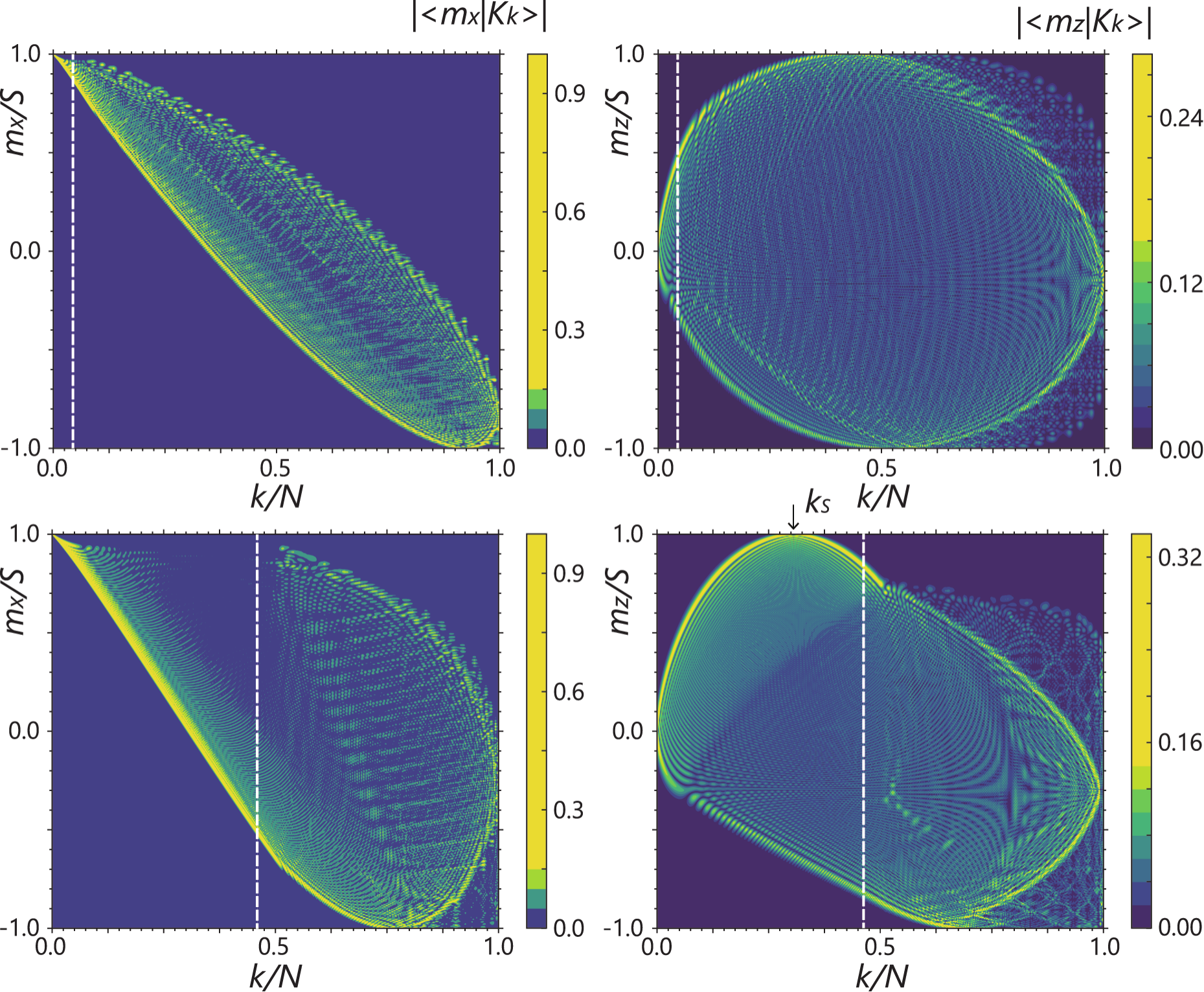}
\caption{
The left panels show $|{}_x\langle m_x|K_k\rangle|$ and 
the right panels $|{}_z\langle m_z|K_k\rangle|$.
We take $(h,g)=(0.5,3.0)$ for the upper panels and 
$(h,g)=(0.5,0.5)$ for the lower panels.
The system size is $N=400$.
The white dashed lines denote the maximum value of the complexity
taken from Fig.~\ref{fig-ks}.
See text for $k_S$.
}
\label{fig-vec}
\end{figure}

The DQPT for the survival probability can be distinguished from that for 
the order parameter~\cite{Marino22, Bento24}.
In this study, we apply the longitudinal field $h$ that breaks 
spin-reflection symmetry.
As a result, it is expected that no sharp transition is observed for 
the magnetization.
We calculate the expectation values of the spin operators $S^z$ and $S^x$
at each time and the result is plotted in Fig.~\ref{fig-spin}.
We observe regular oscillations at large $g$ and 
decaying oscillations at small $g$.
The oscillation period coincides with that of $K(t)$.
The expectation value $\langle \psi(t)|S^z|\psi(t)\rangle$ 
is locally-maximized at the DQPT points
and $\langle \psi(t)|S^x|\psi(t)\rangle$ is locally-minimized.

We also plot in Fig.~\ref{fig-spin} the expectations with respect to 
the ground state of the Hamiltonian.
We find that each time average of $\langle \psi(t)|S^{z,x}|\psi(t)\rangle$ 
is close to the corresponding ground-state expectation at large $g$, 
and deviates from that at small $g$.
In the left panels of Fig.~\ref{fig-spinav}, we compare 
each of the time averages 
$\overline{S}^{z,x}=\int_0^t ds\,\langle \psi(s)|S^{z,x}|\psi(s)\rangle/t$
for a large $t$ to the ground-state expectation.
We see that they are close with each other at large $g$ 
and deviate significantly at small $g$.
The deviation starts around the DQPT point but 
the change is smooth as a function of $g$.
At small $g$, we observe a peak of $\overline{S}^z$, 
which is contrasted to 
the monotonic change of the ground-state expectation.
We find that this peak is related to 
the similar structure of $b_k$ in Fig.~\ref{fig-ab}.
In the range $0<k\le\max_t K(t)$, $b_k$ for a small $g$ has a peak 
at an intermediate value 
and the appearance of the peak corresponds to 
the nonmonotonic behavior of $\overline{S}^z$,
as we show in the right panel of Fig.~\ref{fig-spinav}.

Although the dynamical properties of the system can be understood 
only from the Lanczos coefficients,
it is instructive to see the Krylov basis $|K_k\rangle$.
The left panels of Fig.~\ref{fig-vec} show $|{}_x\langle m_x|K_k\rangle|$.
The initial state $|K_0\rangle$ is localized at $m_x=S$ and 
$|K_k\rangle$ involves smaller $m_x$ contributions as we increase $k$.
The spreading is almost linear
and the basis state reaches $m_x=-S$ at some point smaller than $k=N$.
Since the original Hamiltonian involves
the next-nearest-neighbor hopping in the $x$-basis, it is naively expected that 
the basis state reaches the minimum point $m_x=-S$ at $k=N/2$.
However, the spreading is disturbed by the presence of the other contributions.
We note that the spreading is maximized only 
when there are no other contributions~\cite{Takahashi25}.

The $S^z$-eigenstate-basis distribution of $|K_k\rangle$ 
in the right panels of Fig.~\ref{fig-vec} shows a more complicated behavior.
The initial state in the $z$-basis is written as Eq.~(\ref{psi0z}).
Applying the Krylov expansion, 
we see that the distribution spreads over both the positive and negative directions.
The spreading in the positive direction is faster than that in the negative direction.
The spreading front of the former 
reaches the maximum value $m_z=S$ at a point $k=k_S$.
We display the location of $k_S$ in Fig.~\ref{fig-vec}.
This value of $k_S$ represents the peak of $b_k$ 
in the first domain of the two-block structure.
The decreasing of $b_k$ at $k>k_S$ in the first domain is interpreted as 
a saturation of the state basis.
As a result, we observe a nonmonotonic behavior of $\overline{S}^z$ 
in Fig~\ref{fig-spinav}.

\section{Metastable state}
\label{sec:meta}

\begin{figure}[t]
\centering\includegraphics[width=1.\columnwidth]{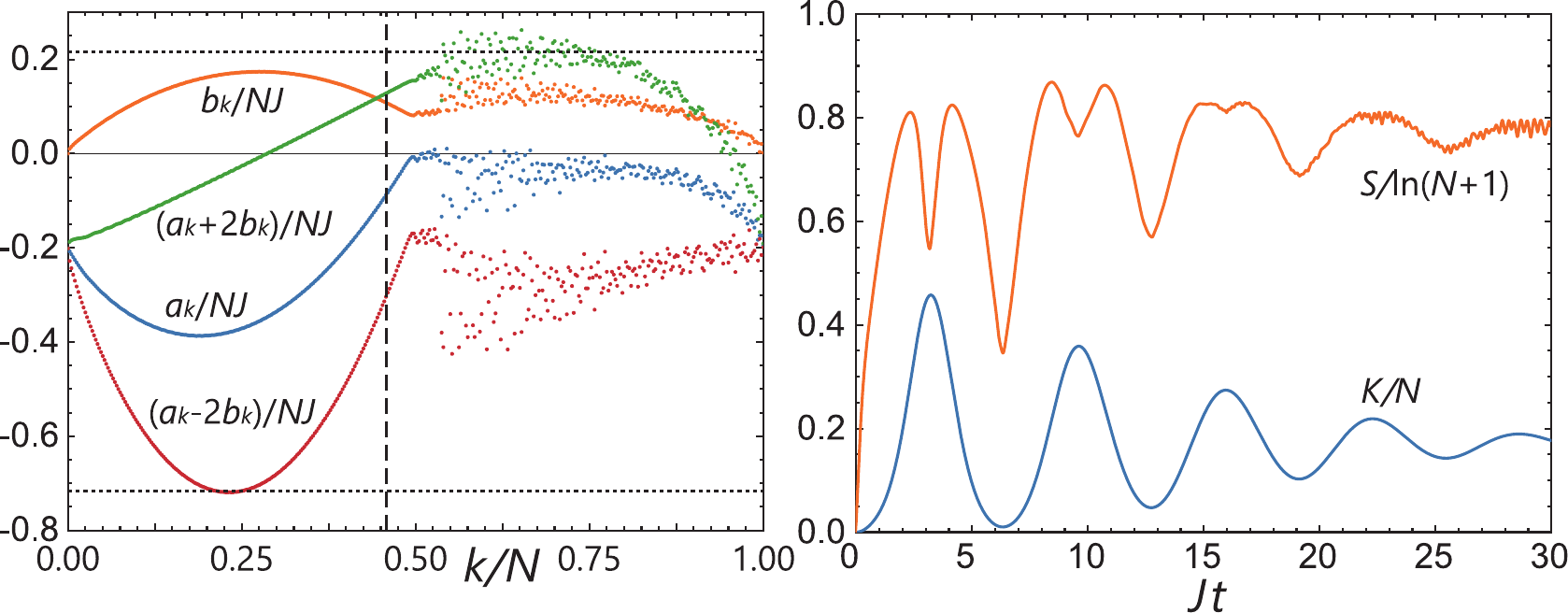}
\caption{
The left panel shows the Lanczos coefficients, and 
the right panel shows the complexity and the entropy.
We take $N=400$ and $(h,g)=(0.2,0.2)$.
}
\label{fig-meta1}
\end{figure}
\begin{figure}[t]
\centering\includegraphics[width=1.\columnwidth]{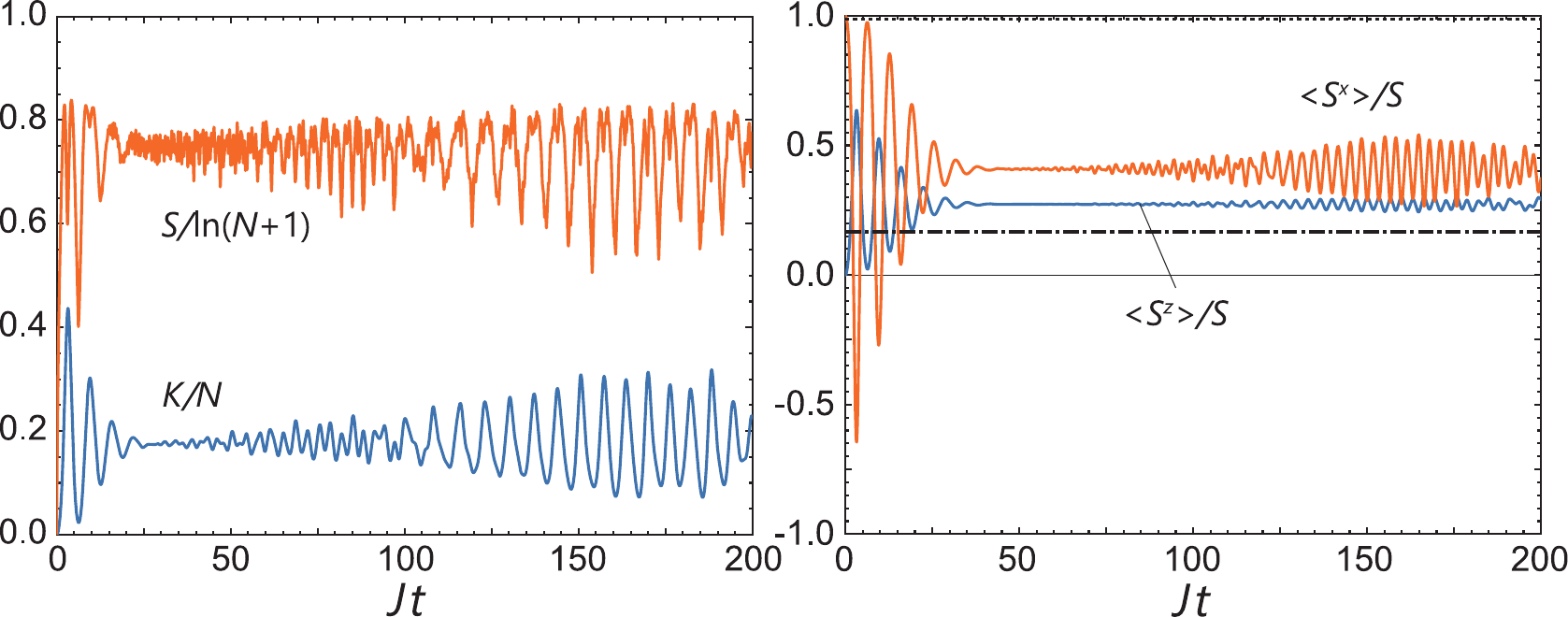}
\caption{
Long-time behavior of the complexity, the entropy, 
and the spin expectations for $N=200$ and $(h,g)=(0.2,0.2)$.
}
\label{fig-meta2}
\end{figure}
\begin{figure}[t]
\centering\includegraphics[width=1.\columnwidth]{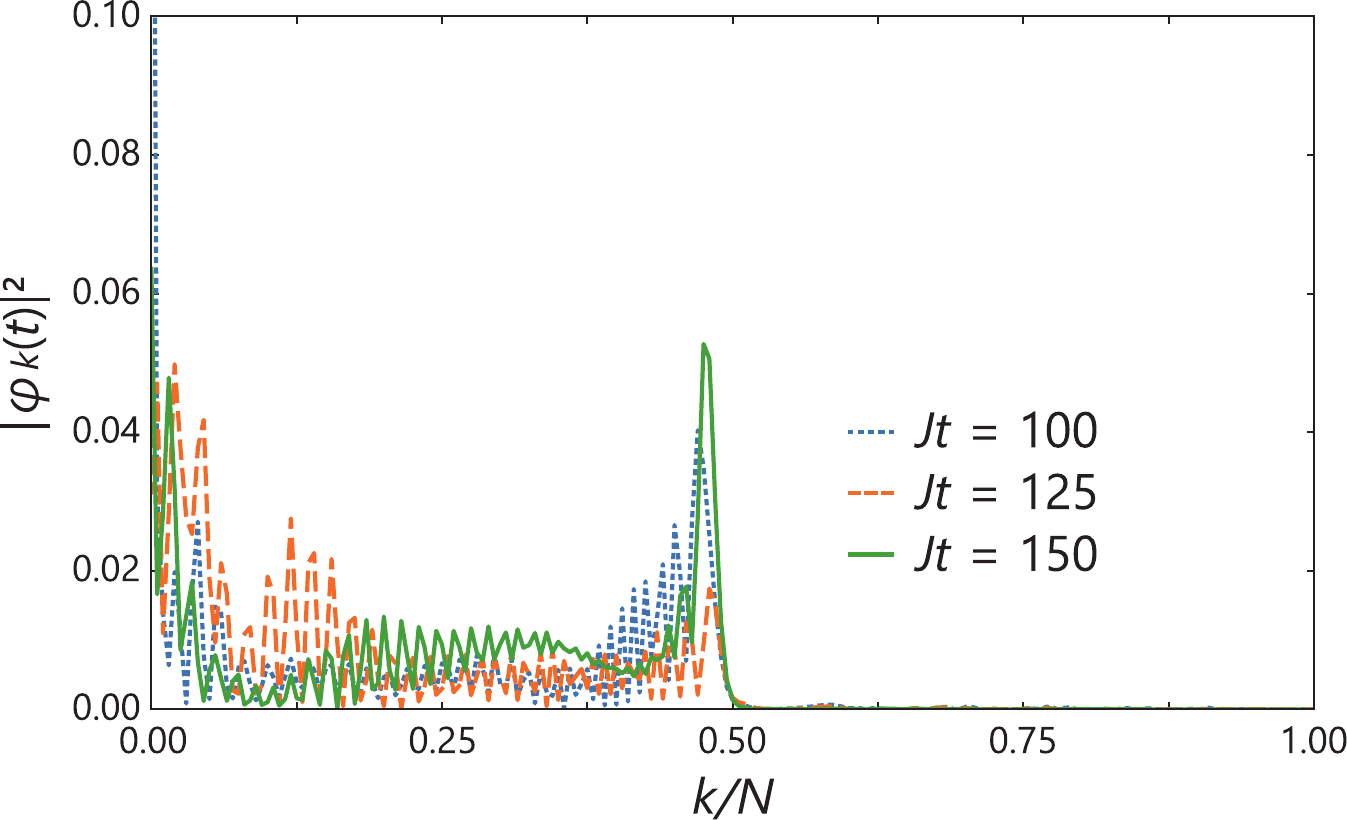}
\caption{
Probability distributions in Krylov space at large $t$.
We take $N=200$ and $(h,g)=(0.2,0.2)$.
}
\label{fig-kdist-meta}
\end{figure}

In this study, we have considered the case where the bias field $h$ is present.
This is because the DQPT is controllable by the field, 
as we see in Eq.~(\ref{rate-exact}). 
Taking $h\to 0$ is not meaningful in that equation.
Numerically, taking a finite $h$ basically gives a stable result due to the unique ground state.

The introduction of the field induces a metastable state.
We need to study how it affects the DQPT.
Generally, in nonequilibrium systems, the metastable state affects the dynamical behavior.
However, the DQPT is obtained only in the thermodynamic limit $N\to\infty$, 
which implies that the metastable state does not play a crucial role for the DQPT.

Figure \ref{fig-pd} shows that the metastable state exists 
when both of $h$ and $g$ are small.
We study the parameter range where the metastable state exists 
and show the results of the Lanczos coefficients, the complexity, and the entropy 
at $(h,g)=(0.2,0.2)$ in Fig .~\ref{fig-meta1}.
Their plots indicate that our picture is basically unchanged.
The Lanczos coefficients show a two-block structure, and 
the wave function extends in the first block.
The DQPTs are found at the reflection points.
At each point, the complexity does not show any singularity and the entropy has a dip.

Taking a closer look of the result gives a different picture.
Although the behavior for small $t$ is basically the same as before, 
it is not for large $t$.
In Fig.~\ref{fig-meta2}, we show the complexity, the entropy, 
and the spin expectations up to a larger value of $t$ with a smaller value of $N$.
They show unstable oscillations at large $t$, which are not observed 
in the absence of the metastable state.
In the left panel of Fig.~\ref{fig-meta1}, 
we observe a decreasing of $a_k$ in the second block at large $k$
giving rise to local minimum points.
Since $a_k-2b_k$ is interpreted as a local potential, 
the emergence of the local minimum
implies the presence of a metastable state in the Krylov space.

The maximum value of the complexity is within the first domain  
as we see in the vertical line in the left panel of Fig.~\ref{fig-meta1}.
In the present case, we need to see not only the average value but also the distribution.
Figure~\ref{fig-kdist-meta} shows that 
the probability for the state to reach the second domain takes a small but nonzero value.
Thus, we can find a Krylov picture of the metastable state.

\section{Dimensionality reduction}
\label{sec:dim}

\begin{figure}[t]
\centering\includegraphics[width=1.\columnwidth]{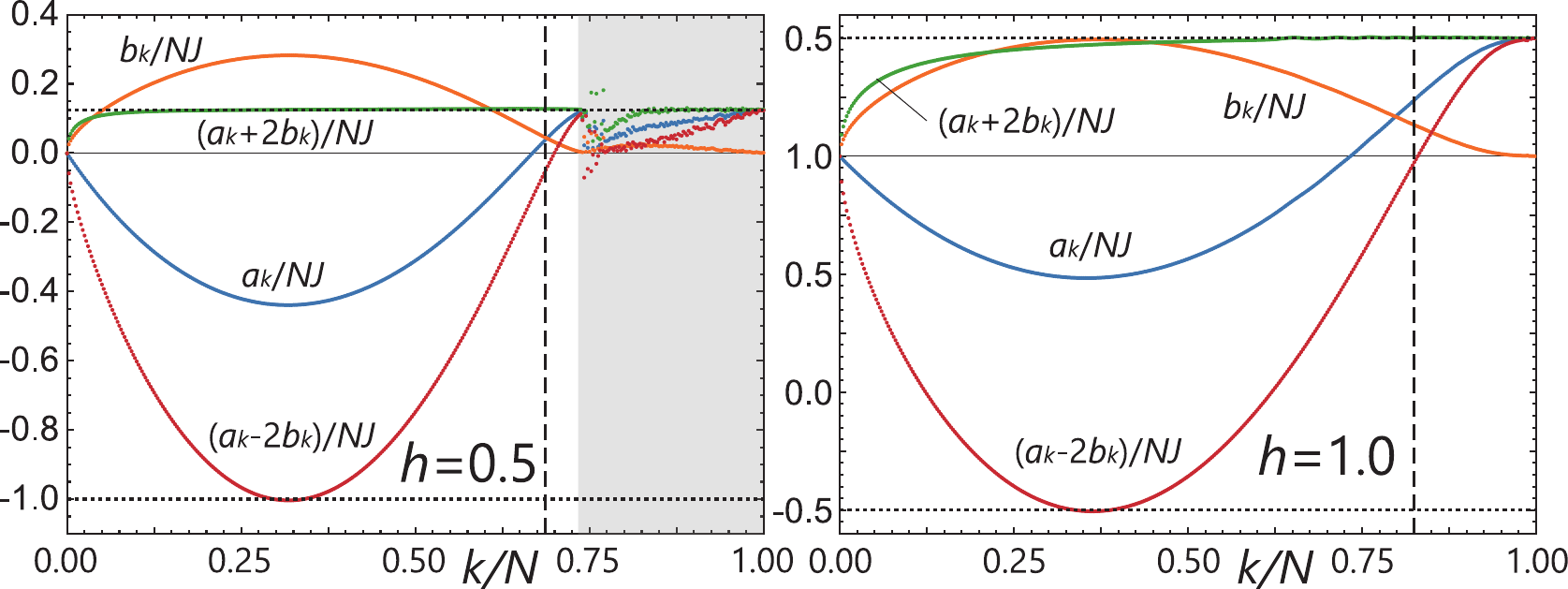}
\caption{
The Lanczos coefficients for $N=400$, $g=0$, $h=0.5$ (left panel) 
and $h=1.0$ (right).
In the left panel, $b_k$ takes a small value at some point $d$ 
with $d<N$ and the shaded domain is discarded.
}
\label{fig-ab-g0}
\end{figure}
\begin{figure}[t]
\centering\includegraphics[width=1.\columnwidth]{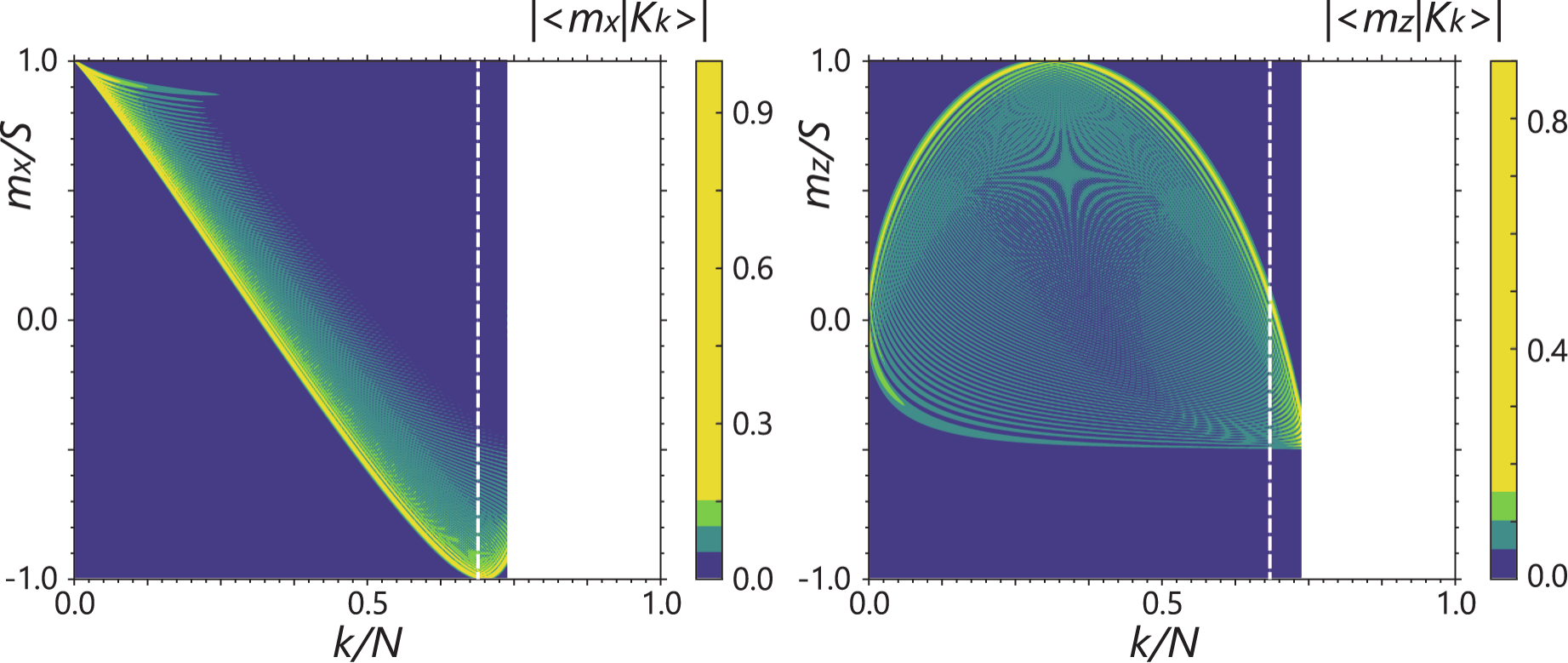}
\caption{
The spin-eigenstate distributions of the Krylov basis.
We take $N=400$ and $(h,g)=(0.5, 0.0)$.
The white dashed lines denote the maximum value of the complexity.
}
\label{fig-vec-g0}
\end{figure}
\begin{figure}[t]
\centering\includegraphics[width=1.\columnwidth]{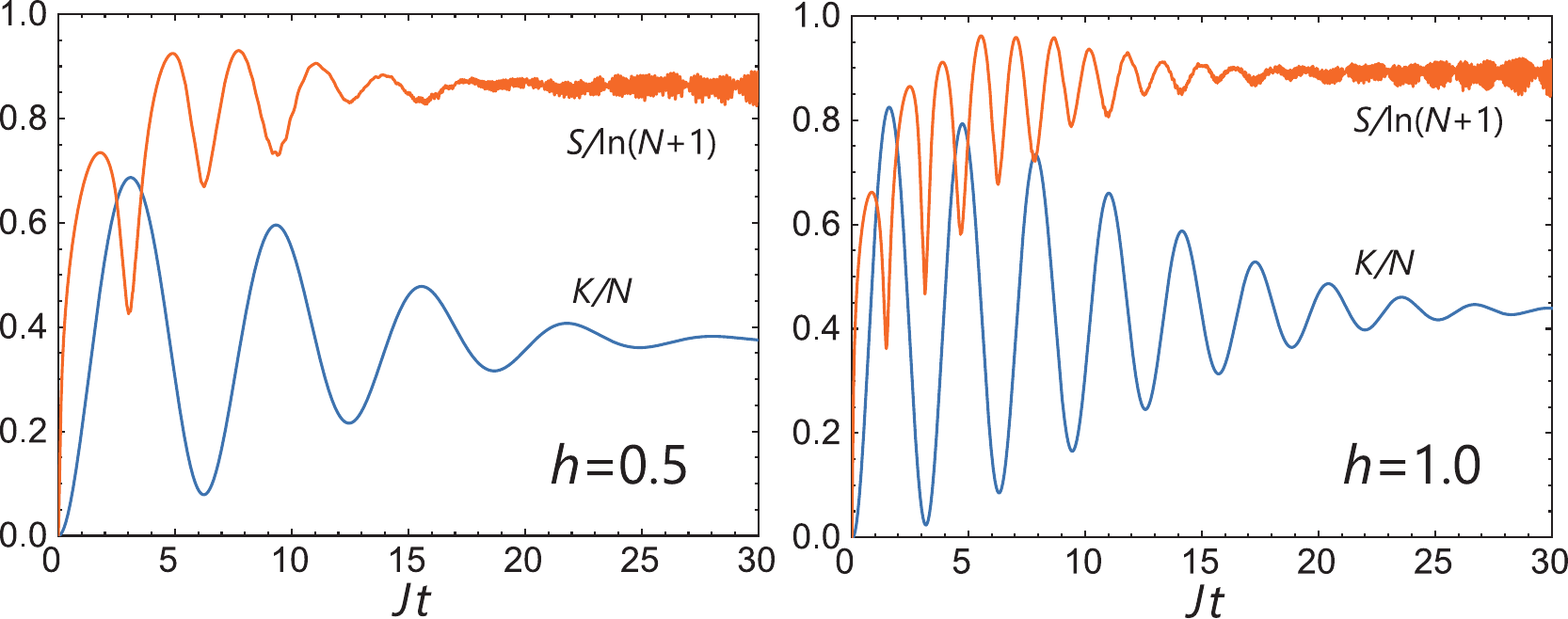}
\caption{
The left panel shows the complexity $K(t)$ and the entropy $S(t)$
at $(h,g)=(0.5,0.0)$, 
and the right panel show those at $(h,g)=(1.0,0.0)$.
We take $N=400$.
}
\label{fig-ks-g0}
\end{figure}

We have treated several cases where the parameters in the Hamiltonian, $h$ and $g$, 
take positive values.
While the bias field $h$ controls the $Z_2$ symmetry of the Hamiltonian,  
the transverse field $g$ controls quantum fluctuation effects induced by the Hamiltonian.
Since the initial state must be different from an eigenstate of the Hamiltonian, 
we may consider a diagonal form of the Hamiltonian to observe the DQPT. 

We discuss the special case at $g=0$ where the Hamiltonian only involves the $S^z$ operator. 
Equation~(\ref{rate-exact}) shows that we can find the DQPT even in that case.
In the $z$-basis representation, 
each of the components evolves independently
from the initial Gaussian distribution in Eq.~(\ref{psi0z}).

We show the result at $g=0$ in 
Figs.~\ref{fig-ab-g0}, \ref{fig-vec-g0}, and \ref{fig-ks-g0}.
As we see in Fig.~\ref{fig-ab-g0}, when $h$ is small, 
$b_k$ takes a very small value at a point smaller than $N$.
This implies that the Krylov dimension is smaller than the Hilbert space dimension.
In principle, the Krylov algorithm halts when we have $b_k=0$.
However, numerical calculations never give the exact value of zero and 
we obtain artificial sequences of the Lanczos coefficients until $k=N$.
In the previous cases such as Figs.~\ref{fig-ab} and \ref{fig-meta1},
$b_k$ takes a very small value only around the end point $k\sim N$.
The result in Fig.~\ref{fig-ab-g0} is very different from the previous results.
In addition, it would be hard to imagine some meaning from 
the behavior in the shaded domain in the left panel of Fig.~\ref{fig-ab-g0}.
Thus, the Krylov dimension at $g=0$ is roughly estimated as
\be
 d(h)\sim\min \left(\frac{1+h}{2}N,N\right).
\ee

Figure \ref{fig-vec-g0} represents the distributions of $|K_k\rangle$
with respect to the spin-eigenstate basis $|m_x\rangle_x$ and $|m_z\rangle_z$.
The spreading in $x$-basis space is almost linear and the basis vector 
reaches the eigenstate $m_x=-S$.
On the other hand, in the $z$-basis case, 
the spreading in the negative direction is suppressed significantly,
which is consistent with the reduction of the Krylov dimension.

The absence of the second block in the Lanczos coefficients implies 
that the effect of the metastable state is negligible.
We do not observe unstable large fluctuations in Fig.~\ref{fig-ks-g0}.
This is due to the simple form of the Hamiltonian.
In the case of $g=0$, the Hamiltonian only contains the $S^z$ operator
and tunneling effects due to quantum fluctuations are absent.

\section{Summary and discussions}
\label{sec:conc}

We have discussed real-time evolutions of quantum states 
for the fully-connected spin model. 
Applying the Krylov algorithm, 
we find that quenched dynamics is described by the spreading in Krylov space. 
The dynamical singularities can occur when the state is reflected 
from a potential barrier in Krylov space.

We found a two-block structure in the Lanczos coefficients.
In contrast to the stable behavior in the first block, 
the second block shows unstable fluctuations.
We however have concluded that the structural change is not related to the DQPT,
because the Krylov wave function basically extends over the first block.
Similar structural changes of the Lanczos coefficients and 
unstable fluctuations in the higher-order components 
can be seen in a broad range of systems~\cite{Nandy25}.
In the present case, the two-block structure may be related 
to the presence of the bias field $h$.
In the absence of $h$, the distributions in Fig.~\ref{fig-vec} will be symmetric 
to prevent nontrivial interference effects.

Since the Lanczos coefficients are related to the difference of the eigenvalues
of the original Hamiltonian, the structural change reminds us 
the excited state quantum phase transition~\cite{Cejnar06,Cejnar21}.
We however confirmed that the diagonalization of the tridiagonal matrix 
does not show any clear structural change.

As we have discussed, 
the DQPT basically appears in the logarithm of the survival amplitude
and it is not simple to find it in the statistical quantities like 
the complexity and the entropy requiring the sum over all components.
They may not be ideal quantities to find the DQPT.
In spite of this result, 
we can predict many dynamical properties of the system 
such as the maximum value of the complexity, 
the existence of metastable states, 
and the nonmonotonic behavior of the order parameter.
These properties can be understood without calculating the time evolution.

Our findings suggest that the Krylov subspace approach provides 
a powerful framework for understanding quantum dynamics beyond conventional methods. 
This perspective not only offers computational advantages but also provides 
deeper physical insights into DQPTs and non-equilibrium phenomena. 
Future work could extend this approach to more complex systems and 
explore connections between the Krylov space structure and 
other quantum information metrics.

\section*{Acknowledgements}
The author is grateful to Adolfo del Campo and Pratik Nandy for useful discussions.
The author acknowledges the financial supports 
from the Luxembourg National Research Fund (FNR Grant No. 16434093)
and from JSPS KAKENHI Grant No. JP24K00547. 
This project has also received funding from 
the QuantERA II Joint Programme with co-funding 
from the European Union’s Horizon 2020 research and innovation programme.


\appendix
\section{Derivation of Eq.~(\ref{slope})}

We derive Eq.~(\ref{slope}).
The initial Krylov basis is set to 
the initial state of the time evolution $|K_0\rangle=|S\rangle_x$.
Applying the Hamiltonian, we obtain 
\be
 && H|K_0\rangle = -\left(Ng+\frac{1}{2}\right)J|S\rangle_x \no\\
 && 
 -\sqrt{N}Jh|S-1\rangle_x-\sqrt{\frac{N-1}{2N}}J|S-2\rangle_x.
 \label{HK0}
\ee
The first term gives the zeroth-order diagonal component $a_0$ as 
\be
 a_0= -\left(Ng+\frac{1}{2}\right)J,
\ee
and the second line is equal to $|K_1\rangle b_1$.
The normalization of $|K_1\rangle$ gives 
\be
 b_1 = \sqrt{Nh^2+\frac{N-1}{2N}}J.
\ee
The first diagonal component $a_1$ is obtained from 
$a_1=\langle K_1|H|K_1\rangle$.
Using the second line of Eq.~(\ref{HK0}), we obtain 
\be
 && \frac{a_1 b_1^2}{J^3} = Nh^2\left(-Ng-2g-\frac{3}{2}+\frac{1}{N}\right)
 \no\\ &&
 +2\sqrt{\frac{N-1}{2}}h\left(-\sqrt{2(N-1)}h\right)
 \no\\ && 
 +\frac{N-1}{2N}\left(-Ng+4g-\frac{5}{2}+\frac{4}{N}\right).
\ee
This gives 
\be
 a_1= -\left(Ng-2g+\frac{7}{2}+O(N^{-1})\right)J,
\ee
and Eq.~(\ref{slope}).

\bibliography{krylov-lmg}

\end{document}